\documentclass[twocolumn,prb,aps,showpacs,amsmath,amssymb]{revtex4}

\usepackage{graphicx}
\usepackage{dcolumn}
\usepackage{bm}

\begin{document}

\title{Kondo effect in interacting nanoscopic systems: Keldysh field integral theory}

\author{Sergey Smirnov}
\email{sergey.smirnov@physik.uni-regensburg.de}
\author{Milena Grifoni}
\affiliation{Institut f\"ur Theoretische Physik, Universit\"at Regensburg,
  D-93040 Regensburg, Germany}

\date{\today}

\begin{abstract}
Kondo physics in nonequilibrium interacting nanoscale devices is an attractive
fundamental many-particle phenomenon with a rich potential for applications.
Due to enormous complexity its clear and flexible theory is still highly
desirable. We develop a physically transparent analytical theory capable to
correctly describe the Kondo effect in strongly interacting systems at
temperatures close to and above the Kondo temperature. We derive a nonequilibrium
Keldysh field theory valid for a system with any finite electron-electron
interaction which is much stronger than the coupling of the system to contacts.
Finite electron-electron interactions are treated involving as many slave-boson
degrees of freedom as one needs for a concrete many-body system. In a small
vicinity of the zero slave-bosonic field configuration weak slave-bosonic
oscillations, induced by the dot-contacts tunneling, are described by an effective
Keldysh action quadratic in the slave-bosonic fields. For clarity the theory is
presented for the single impurity Anderson model but the construction of the
Keldysh field integral is universal and applicable to systems with more complex
many-body spectra.
\end{abstract}

\pacs{72.15.Qm, 73.63.-b, 72.10.Fk}

\maketitle

\section{Introduction}\label{intro}
With the emergence of technological feasibility in manufacturing small
artificial systems with few electrons and using single molecules as electronic
devices, quantum nonequilibrium has taken a new wind lasting already for more than
two decades. On one side the diminutiveness of these nanoscale systems
facilitates their fast passage to a nonequilibrium state already at tiny external
voltages which calls for more comprehensive nonequilibrium theories beyond linear
response. On the other side at these scales electron-electron correlations
start playing a significant role leading to phenomena which are fundamentally
absent in the noninteracting counterparts. A remarkable representative of such
nonequilibrium interacting phenomena is the Kondo effect
\cite{Ralph_1994,Goldhaber-Gordon_1998} revealing itself in the differential
conductance as a zero-bias anomaly having a universal temperature dependence
\cite{Goldhaber-Gordon_1998a,Grobis_2008} with the energy scale known as the
Kondo temperature, $T_\text{K}$.

The Kondo physics turns out to be highly complex. In particular, in its
temperature dependence it has two regimes, weak and strong coupling ones, with
the crossover taking place in the vicinity of the Kondo temperature. This has
given rise to the development of two types of analytical theories valid either
deep below or much above $T_\text{K}$. In the construction of such theories the
concept of a slave-boson has played an extremely fruitful role and it has been
used both for the first \cite{Aguado_2000,Lopez_2003,Ratiani_2009} and second
\cite{Sivan_1996} types of analytical theories. The Kondo effect is governed by
infrared physics. A powerful method capturing this behavior is the renormalization
group method applied to the Kondo model, both in its analytical \cite{Anderson_1970}
and numerical \cite{Wilson_1975} implementations. The analytical RG theory has been
generalized \cite{Rosch_2003,Paaske_2004} to nonequilibrium and provides a
controlled theory at bias voltages $eV>kT_\text{K}$.

Numerical approaches have certain advantages.  For example, the numerical
renormalization group theory \cite{Costi_1994} can cover both the weak and
strong coupling regimes. Unfortunately, its generalization to nonequilibrium
is still problematic. Another numerical method, the noncrossing approximation
(NCA) \cite{Wingreen_1994,Haule_2001}, is able to deal with nonequilibrium
situations but it gets less reliable below $T_\text{K}$. However, the main
disadvantage of all numerical techniques is that they do not provide a clear
picture of a physical phenomenon and, therefore, analytical theories have higher
fundamental priority.

The aim of the present study is to demonstrate that the field-theoretic
approach \cite{Smirnov_2011} based on the Keldysh field integral in the
slave-bosonic representation is a powerful tool in the development of
analytical theories for the Kondo effect in nanoscale systems and that the
derivation of the effective Keldysh action represents a universal,
straightforward and physically clear way to possible theoretical
generalizations inspired by modern experiments. In addition to the technical
flexibility this field-theoretic framework in its original simplest
formulation is already close to the crossover between the weak and strong
coupling regimes: it is valid for temperatures $T\gtrsim T_\text{K}$. Therefore,
after a proper generalization it may penetrate the strong coupling regime and
become an analytical theory of both the first and second type. 

To be specific, here we will develop a generalization to a particularly
complicated and at the same time extremely important aspect related to a
strong but finite electron-electron interaction in a quantum dot (QD). Up to
now many analytical theories (which are mainly of the first type) have only
been able to describe the Kondo physics in QDs with either rather weak
\cite{Muehlbacher_2011} or infinitely strong
\cite{Sivan_1996,Lopez_2003,Ratiani_2009} electron-electron
interactions. Thus, an analytical theory which could be placed between these
two extreme cases is in great demand, especially, if this theory is of the
second type which is more interesting for applications. It is our goal here to
present such a theory.

The paper is organized as follows. In Section \ref{hsb} we formulate the problem
and transform it into a slave-bosonic representation. The effective Keldysh
action is derived in Section \ref{eka}. It is used to derive the tunneling density
of states in Section \ref{tdosr}. With Section \ref{concl} we conclude.

\section{Hamiltonian and slave-bosons}\label{hsb}
We will show a nonperturbative, both in the QD-contacts coupling and
electron-electron interaction, field-theoretic construction for the case of
the single impurity Anderson model (SIAM) of an interacting QD,
\begin{equation}
\hat{H}_\text{d}=\sum_\sigma\epsilon_\text{d}\hat{n}_{\text{d},\sigma}+U\hat{n}_{\text{d},\uparrow}\hat{n}_{\text{d},\downarrow},
\label{SIAM}
\end{equation}
where $\epsilon_\text{d}$ is the single-particle electronic
($\sigma=\uparrow,\downarrow$) energy level of the QD, $U$ is the strength of
the electron-electron interaction and
$\hat{n}_{\text{d},\sigma}=d^\dagger_\sigma d_\sigma$ with $d^\dagger_\sigma$,
$d_\sigma$ being the QD electronic creation and annihilation operators.
The Hamiltonian given by Eq. (\ref{SIAM}) is sufficient to show the basic
steps of the theory without technical complications playing no principal
role in the formalism. The generalization to more complicated models is
straightforward as will become clear from the derivation below.

To complete the formulation of the problem one introduces the tunneling Hamiltonian,
\begin{equation}
\hat{H}_\text{T}=\sum_{a\sigma}(c^\dagger_aT_{a\sigma}d_\sigma+d^\dagger_\sigma T^*_{a\sigma}c_a),
\label{tun_ham}
\end{equation}
and the contacts Hamiltonian,
\begin{equation}
\hat{H}_\text{C}=\sum_a\epsilon_ac^\dagger_ac_a.
\label{cont_ham}
\end{equation}
In Eqs. (\ref{tun_ham}) and (\ref{cont_ham}) $c^\dagger_a$ ($c_a$) are the contacts
electronic creation (annihilation) operators with respect to the state $a$, which
is a quantum electronic state in the left (L) or the right (R) contact. The quantity
$T_{a\sigma}$ in Eq. (\ref{tun_ham}) is the probability amplitude for the tunneling
event from the QD state $\sigma$ to the contact state $a$, and $\epsilon_a$ in
Eq. (\ref{cont_ham}) is the contact electronic single-particle energy corresponding
to the state $a$.

Before constructing the Keldysh field theory we rewrite the QD Hamiltonian,
Eq. (\ref{SIAM}), and the tunneling Hamiltonian, Eq. (\ref{tun_ham}), using
slave-bosons and new fermions replacing the original electrons of the QD. To do
this we will employ the approach developed for magnetic impurities by Barnes \cite{Barnes_1976}.
It was also used later for superconducting compounds by Zou and Anderson \cite{Zou_1988}.
This approach represents a natural extension of the infinite-$U$ slave-boson approach
\cite{Coleman_1984}. An alternative finite-$U$ slave-boson method was proposed by Kotliar
and Ruckenstein \cite{Kotliar_1986}. The essential difference between the two methods is
that the latter one, instead of new fermions, uses magnetic bosons. However, as mentioned in
Ref. \onlinecite{Kotliar_1986}, it does not matter how one introduces slave-particles if the
corresponding constraints are taken into account exactly: the final results must be independent
of the slave-particle scheme.

To apply the Barnes and Zou and Anderson technique one exploits the knowledge of the
many-particle spectrum of the SIAM Hamiltonian, Eq. (\ref{SIAM}): $|0\rangle$,
$\epsilon_0=0$; $|\uparrow\rangle$, $|\downarrow\rangle$,
$\epsilon_\uparrow=\epsilon_\downarrow=\epsilon_\text{d}$; $|\uparrow\downarrow\rangle$,
$\epsilon_{\uparrow\downarrow}=2\epsilon_\text{d}+U$. Introducing the slave-bosonic
creation and annihilation operators $e^\dagger$, $e$ for the empty ($|0\rangle$),
$d^\dagger$, $d$ for the doubly occupied ($|\uparrow\downarrow\rangle$) QD states
($e$-state and $d$-state for later reference) and new fermionic creation and
annihilation operators $p^\dagger_\sigma$, $p_\sigma$ for the singly occupied
($|\uparrow\rangle$, $|\downarrow\rangle$) QD states, so that
$d^\dagger_\sigma=ep^\dagger_\sigma+\sigma d^\dagger p_{-\sigma}$,
$d_\sigma=e^\dagger p_\sigma+\sigma dp^\dagger_{-\sigma}$, Eqs. (\ref{SIAM}) and
(\ref{tun_ham}) can be rewritten as
\begin{equation}
\hat{H}_\text{d}=\sum_\sigma\epsilon_\text{d}p^\dagger_\sigma p_\sigma+(2\epsilon_\text{d}+U)d^\dagger d,
\label{SIAM_sb}
\end{equation}
\begin{equation}
\begin{split}
&\hat{H}_\text{T}=\sum_{a\sigma}[T_{a\sigma}c^\dagger_a(e^\dagger p_\sigma+\sigma dp^\dagger_{-\sigma})+\\
&+T^*_{a\sigma}(ep^\dagger_\sigma+\sigma d^\dagger p_{-\sigma})c_a].
\end{split}
\label{tun_ham_sb}
\end{equation}

The full Hamiltonian is $\hat{H}=\hat{H}_\text{d}+\hat{H}_\text{C}+\hat{H}_\text{T}$.
Since the physical subspace, the QD Fock space, is generated by the four states,
$|0\rangle$, $|\uparrow\rangle$, $|\downarrow\rangle$, $|\uparrow\downarrow\rangle$,
the total number of the slave-bosons and new fermions,
$\hat{Q}\equiv e^\dagger e+\sum_\sigma p^\dagger_\sigma p_\sigma+d^\dagger d$,  must be equal
to one, $\hat{Q}=\hat{I}$. One takes this constraint into account by replacing the total
Hamiltonian with a different one, $\hat{H}\rightarrow\hat{H}+\mu\hat{Q}$, and taking
the limit $\mu\rightarrow\infty$ at the end of all calculations. In comparison with
mean field theories this projection onto the physical subspace is exact. Therefore, it
is also applicable for high temperatures and is suitable for a development of both the
first and second type analytical theories.

At this point we would like to mention that, as is obvious from above, the applicability
of the Barnes and Zou and Anderson approach is not restricted only by SIAM, Eq. (\ref{SIAM}).
It is much more universal and applicable to an arbitrary many-particle system for which its
many-body spectrum has been found. Then one can, in full analogy with above, introduce
slave-bosons for the states with the even particle numbers and new fermions for the states
with the odd particle numbers.

\section{Effective Keldysh action}\label{eka}
Using Eqs. (\ref{cont_ham}), (\ref{SIAM_sb}) and (\ref{tun_ham_sb}) we develop a
slave-bosonic Keldysh field theory. To this end we construct the Keldysh field
integral \cite{Altland_2010}. The field theory is obtained performing the functional
integration over all the fermionic fields and eventually is formulated through the
effective Keldysh action. In comparison with Ref. \onlinecite{Smirnov_2011} this action depends
now not only on the $e$-state slave-bosonic field, which we denote as $\chi(t)$, but also
on the $d$-state slave-bosonic field, which we denote as $\xi(t)$:
\begin{equation}
\begin{split}
&S_\text{eff}[\chi^\text{cl}(t),\chi^\text{q}(t);\xi^\text{cl}(t),\xi^\text{q}(t)]=
S^{(0)}_{\text{B}e}[\chi^\text{cl}(t),\chi^\text{q}(t)]+\\
&+S^{(0)}_{\text{B}d}[\xi^\text{cl}(t),\xi^\text{q}(t)]\!+\!S_\text{tun}[\chi^\text{cl}(t),\chi^\text{q}(t);\xi^\text{cl}(t),\xi^\text{q}(t)],
\end{split}
\label{eff_act}
\end{equation}
where $\chi^{\text{cl}(\text{q})}(t)$, $\xi^{\text{cl}(\text{q})}(t)$ are the classical (quantum)
components of the slave-bosonic fields, the actions $S^{(0)}_{\text{B}e}$ and $S^{(0)}_{\text{B}d}$
describe the free dynamics of the $e$- and $d$-states, respectively, and the action
$S_\text{tun}$ embodies the complex tunneling dynamics of both the $e$- and $d$-states.

Any QD observable $\hat{O}=\mathcal{F}(d^\dagger_\sigma,d_\sigma)$ admits the Keldysh field
integral representation in terms of the effective Keldysh action Eq. (\ref{eff_act}),
\begin{equation}
\begin{split}
&\langle\hat{O}\rangle(t)=\frac{1}{\mathcal{N}_0}\underset{\mu\rightarrow\infty}{\text{lim}}e^{\beta\mu}\times\\
&\times\int\mathcal{D}[\chi(t),\xi(t)]e^{\frac{i}{\hbar}S_\text{eff}[\chi^\text{cl}(t),\chi^\text{q}(t);\xi^\text{cl}(t),\xi^\text{q}(t)]}\times\\
&\times\mathcal{F}[\chi^\text{cl}(t),\chi^\text{q}(t);\xi^\text{cl}(t),\xi^\text{q}(t)],
\end{split}
\label{observ_kfi}
\end{equation}
where $\mathcal{N}_0$ is the normalization factor (see Ref. \onlinecite{Smirnov_2011}) and $\beta$
is the inverse temperature.

The general structure of $S_\text{tun}$ is the same as in Ref. \onlinecite{Smirnov_2011},
$S_\text{tun}[\chi^\text{cl}(t),\chi^\text{q}(t);\xi^\text{cl}(t),\xi^\text{q}(t)]=
-i\hbar\,\text{tr}\ln[I+\mathcal{T}G^{(0)}]$. However, the matrices $G^{(0)}$ and $\mathcal{T}$
are now different. The free Green's function matrix $G^{(0)}$ has now an explicit spin
dependence,
\begin{equation}
G^{(0)}=
\begin{pmatrix}
\sigma\,G^{(0)}_\text{d}(\sigma t|\sigma't')&0\\
0&\sigma\,G^{(0)}_\text{C}(\bar{a}\sigma t|\bar{a}'\sigma' t')
\end{pmatrix},
\label{G0_matr}
\end{equation}
where $G^{(0)}_\text{d,C}$ are the fermionic Keldysh Green's function matrices for the isolated
QD and contacts. These matrices are diagonal in the spin space. In Eq. (\ref{G0_matr}) we
have picked out the spin index from the state $a$, {\it i.e.}, $a\equiv\{\bar{a},\sigma\}$
and $\bar{a}$ is the state $a$ but the spin. This explicit spin dependence appears because the
mapping $\{d_\sigma^\dagger,d_\sigma\}\rightarrow\{e^\dagger,e;d^\dagger,d;p_\sigma^\dagger,p_\sigma\}$ has
an explicit spin dependence. What is more crucial is that now, due to this spin-dependent mapping,
the tunneling matrix $\mathcal{T}$ is not anymore diagonal in the spin space. It has the same
block structure as in Ref. \onlinecite{Smirnov_2011} but its blocks have a different architecture in the
spin space:
\begin{equation}
M_\text{T}(\bar{a}\sigma t|\sigma't')=\frac{1}{\hbar}\delta(t-t')
\begin{pmatrix}
T_{\bar{a}\uparrow\uparrow}(t)&\tilde{T}_{\bar{a}\uparrow\downarrow}(t)\\
-\tilde{T}^*_{\bar{a}\downarrow\uparrow}(t)&-T^*_{\bar{a}\downarrow\downarrow}(t)
\end{pmatrix},
\label{mt_matr}
\end{equation}
where $T_{\bar{a}\sigma\sigma'}(t)\equiv T_{\bar{a}\sigma\sigma'}\hat{\chi}^\dagger(t)$,
$\tilde{T}_{\bar{a}\sigma\sigma'}(t)\equiv -\sigma'T_{\bar{a}\sigma\,-\sigma'}\hat{\xi}(t)$. Here
$T_{\bar{a}\sigma\sigma'}=\delta_{\sigma\sigma'}T$ which assumes a symmetric coupling to the contacts.
The matrix $\hat{\chi}(t)$ is the same as in Ref. \onlinecite{Smirnov_2011} and the new matrix
$\hat{\xi}(t)$ is
\begin{equation}
\hat{\xi}(t)\equiv\frac{1}{\sqrt{2}}
\begin{pmatrix}
\xi^\text{q}(t)&\xi^\text{cl}(t)\\
\xi^\text{cl}(t)&\xi^\text{q}(t)
\end{pmatrix}.
\label{chi_xi_matr}
\end{equation}

One can see from Eqs. (\ref{mt_matr}) and (\ref{chi_xi_matr}) that the off-diagonal blocks
in the spin space have an anomalous matrix structure in the Keldysh space: the classical
components are the off-diagonal elements while the quantum components are on the main
diagonal. Physically this non-diagonal spin structure and the anomalous structure in the
Keldysh space are the consequence of the QD population fluctuations between one and two
electrons. These fluctuations are now allowed due to the finite value $U$ of the
electron-electron interaction.

At this point it is important to mention that in our field theory there is no need to scan
over a huge family of diagrams in order to find diagrams recovering the symmetry between
virtual transitions into the empty and doubly occupied states. Violation of this symmetry
represents the main problem in extensions \cite{Haule_2001} of NCA for finite-$U$ systems.
In contrast, in our field theory on one side one does not have to battle with the diagrammatic
combinatorics and on the other side one does not need to care about the symmetry violation
since both virtual transitions are equally treated by keeping in the effective Keldysh action
the same powers of both the empty and double occupancy slave-bosonic fields. Below we keep terms
only up to the second order in these fields and it is obvious that the same can be done at any
order. From the diagrammatic perspective the theory could be close in spirit to the equilibrium
imaginary-time approach \cite{Jin_1988} for the infinite-$U$ Anderson model.

As we are going to construct an analytical theory of the second type, we consider a QD
in the Kondo regime formed by weak (see Ref. \onlinecite{Smirnov_2011}) slave-bosonic
oscillations excited by the electronic tunneling between the QD and contacts. In this
case one can expand $S_\text{tun}$ around the zero configurations of the slave-bosonic
fields $\chi(t)$ and $\xi(t)$ up to the second order:
\begin{equation}
\begin{split}
&S_\text{tun}[\chi^\text{cl}(t),\chi^\text{q}(t);\xi^\text{cl}(t),\xi^\text{q}(t)]=\\
&=\frac{\pi\Gamma}{2\hbar}\int dt\int dt'
\begin{pmatrix}
\bar{\chi}^\text{cl}(t)&\bar{\chi}^\text{q}(t)
\end{pmatrix}\times\\
&\times\begin{pmatrix}
0&\Sigma_{\text{B}e}^-(t-t')\\
\Sigma_{\text{B}e}^+(t-t')&\Sigma_{\text{B}e}^\text{K}(t-t')
\end{pmatrix}
\begin{pmatrix}
\chi^\text{cl}(t')\\
\chi^\text{q}(t')
\end{pmatrix}-\\
&-\frac{\pi\Gamma}{2\hbar}\int dt\int dt'
\begin{pmatrix}
\bar{\xi}^\text{cl}(t)&\bar{\xi}^\text{q}(t)
\end{pmatrix}\times\\
&\times\begin{pmatrix}
0&\Sigma_{\text{B}d}^-(t-t')\\
\Sigma_{\text{B}d}^+(t-t')&\Sigma_{\text{B}d}^\text{K}(t-t')
\end{pmatrix}
\begin{pmatrix}
\xi^\text{cl}(t')\\
\xi^\text{q}(t')
\end{pmatrix},
\end{split}
\label{tun_act_exp}
\end{equation}
where $\Sigma_{\text{B}e}^{+,-,\text{K}}$ and $\Sigma_{\text{B}d}^{+,-,\text{K}}$ are the slave-bosonic
self-energies of the $e$- and $d$-states. The self-energies $\Sigma_{\text{B}e}^{+,-,\text{K}}$ have
been derived in Ref. \onlinecite{Smirnov_2011} and the new objects, $\Sigma_{\text{B}d}^{+,-,\text{K}}$,
are
\begin{equation}
\begin{split}
&\Sigma_{\text{B}d}^\pm(t-t')\equiv\frac{i}{2}\sum_x[g^\text{K}_x(t-t')g^\pm_\text{d}(t-t')+\\
&+g^\pm(t-t')g^\text{K}_\text{d}(t-t')],\\
&\Sigma_{\text{B}d}^\text{K}(t-t')\equiv\frac{i}{2}\sum_x\{g^\text{K}_x(t-t')g^\text{K}_\text{d}(t-t')+\\
&\!\!\!+[g^+_\text{d}(t-t')-g^-_\text{d}(t-t')][g^+(t-t')-g^-(t-t')]\},
\end{split}
\label{se_d}
\end{equation}
where $x=\text{L},\text{R}$ and the functions $g^{+,-,\text{K}}_\text{d}$, $g^\text{K}_x$, $g^\pm$ are
the same as in Ref. \onlinecite{Smirnov_2011}. We use the Lorentzian density of states for the contacts,
$\nu_\text{C}(\epsilon)=\nu_\text{C}W^2/(\epsilon^2+W^2)$, and define
$\Gamma\equiv 4\pi\nu_\text{C}|T|^2$.

The advantage of this quadratic model is that it is analytically solvable. However, its
applicability is limited by temperatures $T\gtrsim T_\text{K}$ and QD chemical potentials
$\Gamma\lesssim\mu_0-\epsilon_\text{d}\lesssim U-\Gamma$ which assumes that the theory is valid for
$U\gg\Gamma$. The latter is exactly our primary goal.

Let us say a few words about the structure of the tunneling action. As one can see from Eq.
(\ref{tun_act_exp}), it does not contain terms mixing $\chi(t)$ and $\xi(t)$. In the second order
such terms just do not appear. The physical picture behind this is that the charge excitations
corresponding to the single and double occupancies do not interact. This is valid for large $U$.
However, even for large $U$ the excitations can interact at low temperatures through the Kondo
resonance for $\mu_0$ in the vicinity of the symmetric point, $\epsilon_\text{d}+U/2$. Therefore,
one expects that in this vicinity the quadratic theory is not valid at low temperatures.

\section{Tunneling density of states and results}\label{tdosr}
It is now a straightforward task to calculate observables using Eq. (\ref{tun_act_exp}). The
observable storing the whole universe of the Kondo physics in QDs is the tunneling density of states
(TDOS),
$\nu_\sigma(\epsilon)\equiv -(1/\hbar\pi)\text{Im}[G_{\text{d}\,\sigma\sigma}^+(\epsilon)]$, where
$G_{\text{d}\,\sigma\sigma'}^+(\epsilon)$ is the retarded QD Green's function. Using the functional
integral representation, Eq. (\ref{observ_kfi}), and performing the Gaussian integral we obtain
\begin{equation}
\begin{split}
&\nu_\sigma(\epsilon)=\mathcal{Z}(\epsilon)\biggl\{\frac{1}{[\epsilon_\text{d}-\epsilon+g\Sigma_e^\text{R}(\epsilon)]^2+[g\Sigma_e^\text{I}(\epsilon)]^2}+\\
&+\frac{1}{[\epsilon_\text{d}+U-\epsilon+g\Sigma_d^\text{R}(\epsilon)]^2+[g\Sigma_d^\text{I}(\epsilon)]^2}\biggl\},
\end{split}
\label{tdos}
\end{equation}
where $g\equiv\pi\Gamma/2$. In Eq. (\ref{tdos}) $\Sigma_e^{\text{R}(\text{I})}$ is the real (imaginary)
part of the $e$-state retarded slave-bosonic self-energy (see Ref. \onlinecite{Smirnov_2011}) and
$\Sigma_d^{\text{R}(\text{I})}$ is the real (imaginary) part of the $d$-state retarded slave-bosonic
self-energy, $\Sigma_d^{\text{R}}(\epsilon)=W\epsilon/[\pi(\epsilon^2+W^2)]-\Sigma_e^{\text{R}}(\epsilon)$,
$\Sigma_d^{\text{I}}(\epsilon)=-W^2[2-n_\text{L}(\epsilon)-n_\text{R}(\epsilon)]/[2\pi(\epsilon^2+W^2)]$,
where $n_\text{L}$, $n_\text{R}$ are the Fermi-Dirac distributions of the contacts electrons,
$n_{\text{L}(\text{R})}(\epsilon)=\{\exp[\beta(\epsilon-\mu_0\pm eV/2)]+1\}^{-1}$, characterized by an
external voltage $V$. The function $\mathcal{Z}$ in Eq. (\ref{tdos}) is
\begin{equation}
\begin{split}
&\mathcal{Z}(\epsilon)\equiv\frac{\Gamma}{4\pi}\frac{W^2}{\epsilon^2+W^2}\times\\
&\times\frac{n(\epsilon_\text{d})[1-n_{\text{B}d}(\epsilon_\text{d})]}
{1-n_\text{L}(\epsilon_\text{d})n_\text{R}(\epsilon_\text{d})-n_{\text{B}d}(\epsilon_\text{d})n(\epsilon_\text{d})},
\end{split}
\label{z_func}
\end{equation}
where $n(\epsilon)\equiv n_\text{L}(\epsilon)+n_\text{R}(\epsilon)-2n_\text{L}(\epsilon)n_\text{R}(\epsilon)$,
$n_{\text{B}d}(\epsilon)\equiv\{\exp[\beta(2\epsilon+U-2\mu_0)]+1\}^{-1}$.
\begin{figure}
\includegraphics[width=8.4 cm]{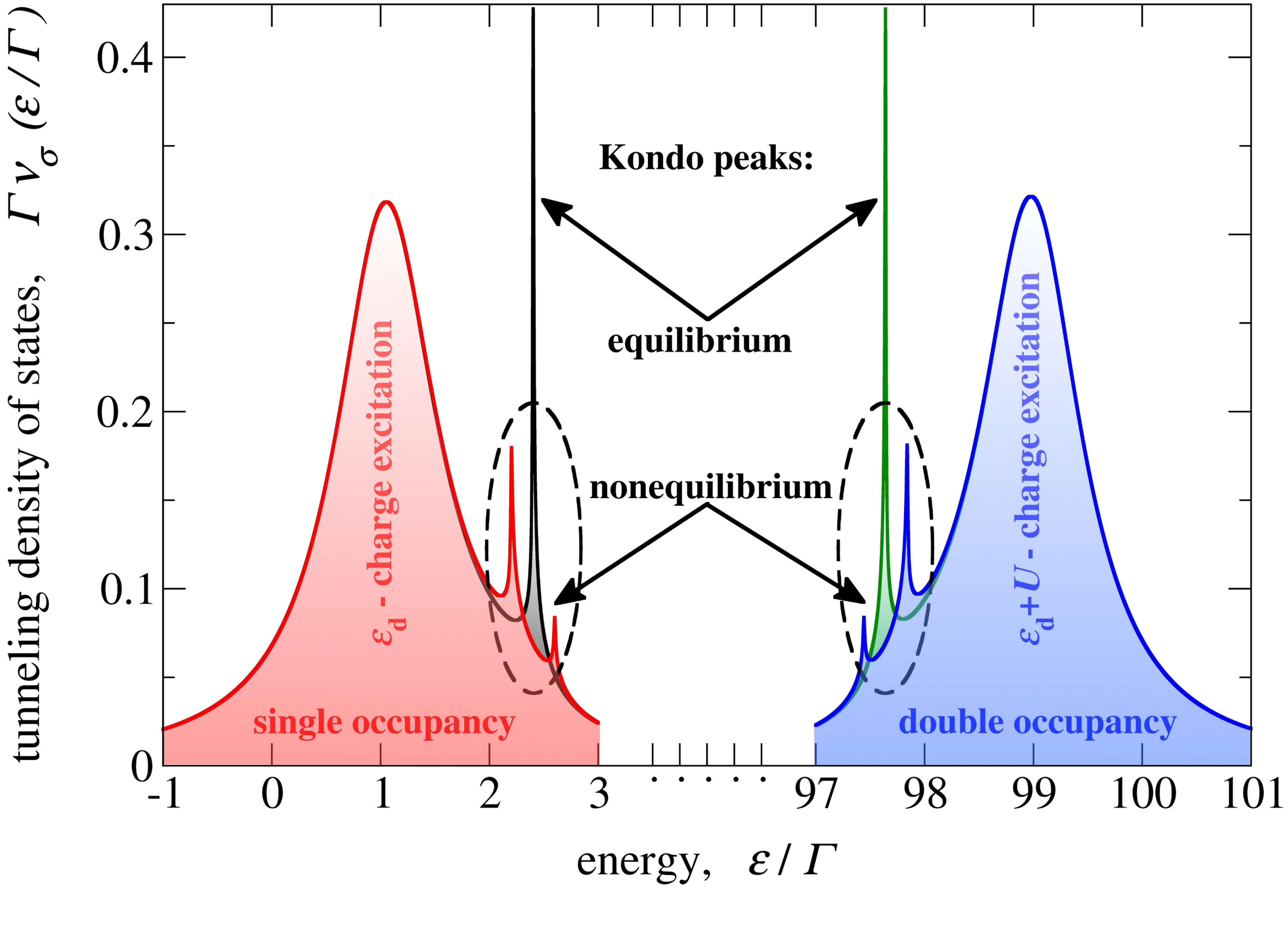}
\caption{\label{figure_1} (Color online) The analytical result, Eq. (\ref{tdos}), for the QD TDOS.
  Here $kT=0.0035\Gamma$, $\epsilon_\text{d}=0$, $\mu_0-\epsilon_\text{d}=2.4\Gamma$ and $97.6\Gamma$,
  $U=100\Gamma$, $W=1000\Gamma$. For nonequilibrium $eV=0.4\Gamma$.}
\end{figure}

The QD TDOS, Eq. (\ref{tdos}), is shown in Fig. \ref{figure_1}. Since $U$ is finite, the TDOS has two
charge excitations, the single and double occupancy excitations close to $\epsilon_\text{d}$ and
$\epsilon_\text{d}+U$, respectively. For a QD in which the strength $U$ of the electron-electron
\begin{figure}
\includegraphics[width=7.83 cm]{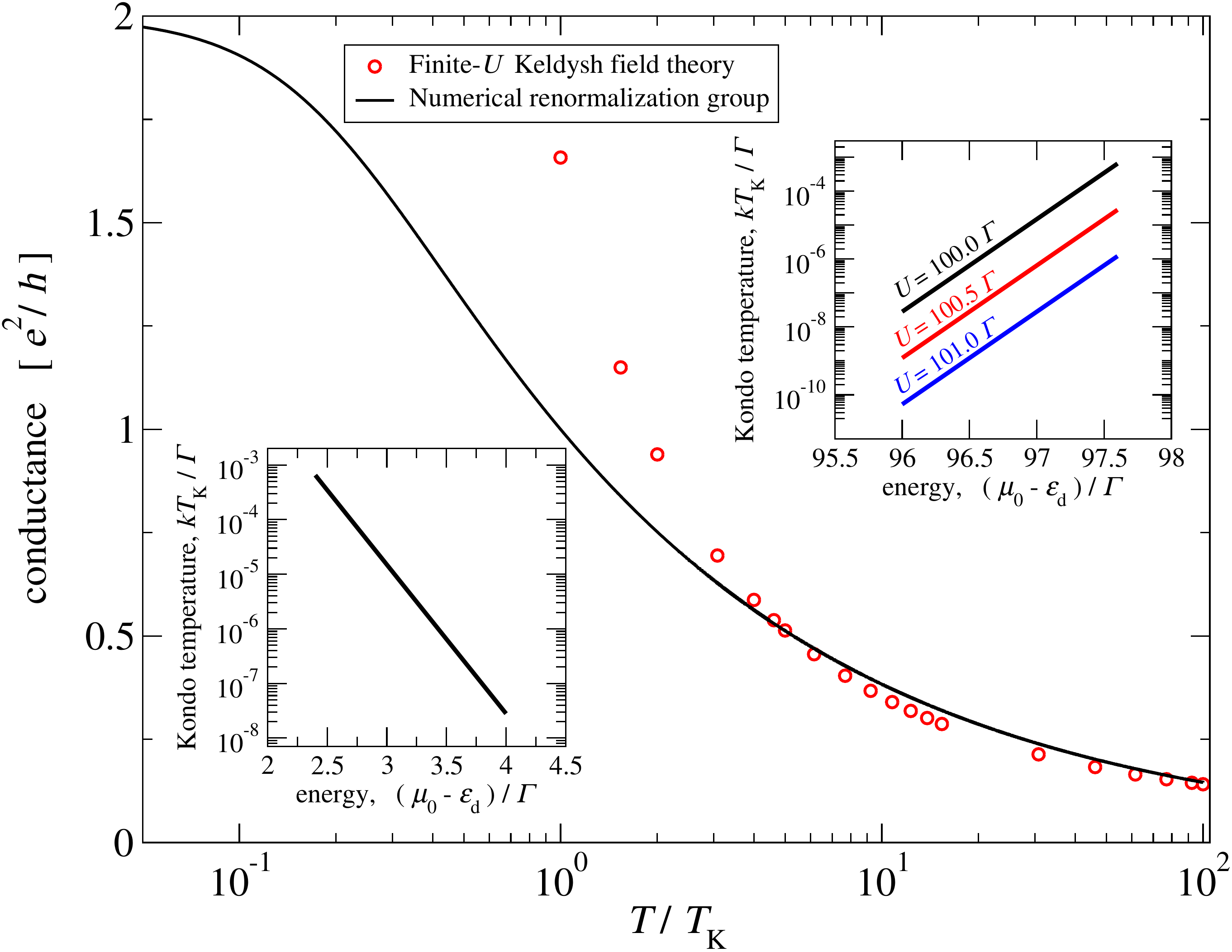}
\caption{\label{figure_2} (Color online) The universal temperature dependence of the conductance.
  Insets show the two qualitatively different regimes in the Kondo temperature dependence on
  $\mu_0-\epsilon_\text{d}$.}
\end{figure}
interaction is finite the behavior of the Kondo resonance as a function of $\mu_0-\epsilon_\text{d}$
is quite different from the limiting case $U=\infty$. For a finite value of $U$ the Kondo peak first
decreases when $\mu_0-\epsilon_\text{d}$ increases up to the symmetric point $U/2$ and after this
point increases again when $\mu_0-\epsilon_\text{d}$ approaches $U$, so that two symmetrically located
Kondo peaks, close to $\epsilon_\text{d}$ and $\epsilon_\text{d}+U$, have the same height as demonstrated
in Fig. \ref{figure_1}. This means that the Kondo temperature $T_\text{K}$ as a function of
$\mu_0-\epsilon_\text{d}$ decreases up to $U/2$ and then increases again. Our Keldysh field theory
correctly predicts this behavior of $T_\text{K}$ obtained from the scaling in the universal temperature
dependence of the conductance as shown in Fig. \ref{figure_2}. This result agrees with the general
expression (see, {\it e.g.}, Ref. \onlinecite{Goldhaber-Gordon_1998a}),
$kT_\text{K}/\Gamma\thicksim(W/\Gamma)\exp\{-2\pi(\mu_0-\epsilon_\text{d})[U-(\mu_0-\epsilon_\text{d})]/\Gamma U\}$.
In Fig. \ref{figure_2} we also compare our theory with the numerical renormalization group theory
\cite{Costi_1994,Goldhaber-Gordon_1998a,Grobis_2008}. This comparison proves that, indeed, our theory
is of the second type and it is reliable for temperatures $T\geqslant 2T_\text{K}$.

\section{Conclusion}\label{concl}
In summary, we have developed an analytical nonequilibrium Keldysh field theory for the Kondo effect
in QDs with finite electron-electron interactions ($U$) which are much stronger than the QD-contacts
coupling ($\Gamma$). The theory is nonperturbative in both $U$ and $\Gamma$ and valid for temperatures
$T\geqslant 2T_\text{K}$. Although, for clarity, it has been presented for the case of SIAM with normal
contacts, the construction of the Keldysh field integral has a universal theoretic scheme applicable to
setups with ferromagnetic or superconducting contacts coupled to any interacting nanoscopic system after
its many-particle spectrum has been found.

The authors thank Rosa L{\'{o}}pez for useful discussions. Support from the DFG SFB 689 is acknowledged.

\end{document}